\documentclass[letter,11pt]{article}

\usepackage{fullpage}
\usepackage{amsfonts,cite}
\usepackage{afterpage,layout,longtable,varioref,verbatim}
\usepackage{ifthen}
\usepackage{enumerate}
\usepackage[T1]{fontenc}
\usepackage[noend]{algorithmic}
\usepackage[latin1]{inputenc}
\usepackage{float}

\floatstyle{ruled}
\newfloat{algorithm}{tbp}{loa}
\floatname{algorithm}{Algorithm}
\usepackage{graphics}
\usepackage{color}
\usepackage{latexsym}
\usepackage{epsf,psfrag}
\usepackage{epsfig}
\usepackage{amssymb}
\usepackage{textcomp}
\usepackage[mathscr]{eucal} 
\usepackage{multirow} 

\usepackage[cmex10]{amsmath}
\usepackage[caption=false]{subfig}
 \usepackage{cases}

\newcommand\R{\mathbb{R}}
\newcommand\E{\mathbb{E}}
\newcommand\Prb{\mathbb{P}}

\newcommand\diam{\operatorname{diam}}

\newcommand\Lm[1]{Lemma\/~\textup{\ref{l:#1}}}
\newcommand\Co[1]{Corollary\/~\textup{\ref{c:#1}}}

\def\Bmsc{{Q}}
\def\Rd{{\mathbb{R}^d}}

\def\SP{{\mathrm{SP}}}
\def\tmst{{\mbox{\tiny MST}}}
\def\Sum{{\mbox{\tiny AGG}}}
\def\Gen{{\mbox{\tiny CLQ}}}

\newcommand{\Vmsc}{\mathscr{V}}

\DeclareMathOperator{\mst}{MST}

\DeclareMathOperator{\dist}{dist}

\DeclareMathOperator{\Var}{Var}

 \def\0{{\bf 0}}

\def\st{{s.t.  }}

\def\nn{\nonumber}

\def\qed{\hfill\hbox{${\vcenter{\vbox{
    \hrule height 0.4pt\hbox{\vrule width 0.4pt height 6pt
    \kern5pt\vrule width 0.4pt}\hrule height 0.4pt}}}$}}

\definecolor{myred}{rgb}{0.3,0.0,0.7}
\definecolor{dkg}{rgb}{0.1,0.7,0.2}
\definecolor{dkb}{rgb}{0.0,0.2,0.8}

\newcommand{\Cmsc}{\mathscr{C}}

\newcommand{\Gmsc}{\mathscr{G}}

\def\bfw{{\mathbf w}}

\def\bfV{{\mathbf V}}

\def\bfY{{\mathbf Y}}

\def\Bc{{\cal B}}
\def\Cc{{\cal C}}

\def\Ec{{\cal E}}

\def\Pc{{\cal P}}

\def\Sc{{\cal S}}

\def\Yc{{\cal Y}}

\def\Ebb{{\mathbb E}}

\def\Nbb{{\mathbb N}}

\newcommand{\bprf}{\begin{myproof}}
\newcommand{\eprf}{\end{myproof}}
\newcommand{\bp}{\begin{psfrags}}
\newcommand{\ep}{\end{psfrags}}
\newcommand{\bl}{\begin{lemma}}
\newcommand{\el}{\end{lemma}}
\newcommand{\bt}{\begin{theorem}}
\newcommand{\et}{\end{theorem}}
\newcommand{\bc}{\begin{center}}
\newcommand{\ec}{\end{center}}
\newcommand{\bi}{\begin{itemize}}
\newcommand{\ei}{\end{itemize}}
\newcommand{\ben}{\begin{enumerate}}
\newcommand{\een}{\end{enumerate}}
\newcommand{\bd}{\begin{definition}}
\newcommand{\ed}{\end{definition}}
\def\beq{\begin{equation}}
\def\eeq{\end{equation}\noindent}
\def\beqn{\begin{eqnarray}}
\def\eeqn{\end{eqnarray} \noindent}
\def\beqnn{  \begin{eqnarray*}}
\def\eeqnn{\end{eqnarray*}  \noindent}
\def\bcase{  \begin{numcases}}
\def\ecase{\end{numcases}   \noindent}
\def\bsbcase{  \begin{subnumcases}}
\def\esbcase{\end{subnumcases}   \noindent}

\def\defeq{{:=}}

\newtheorem{theorem}{Theorem}
\newtheorem{corollary}{Corollary}
\newtheorem{lemma}{Lemma}
\newtheorem{definition}{Definition}

\newtheorem{proposition}{Proposition}

\newenvironment{myproof}{\noindent{\em Proof:} \hspace*{1em}}{
    \hspace*{\fill} $\Box$ }
\newenvironment{proof_of}[1]{\noindent {\em Proof of #1: }}{\hspace*{\fill} $\Box$ }

\newcommand{\matplottc}[1]{               
        \unitlength .45truein
        \begin{center}
        \includegraphics{#1.ps}
        \end{picture}
        \end{center}
}

\def\psfancypar#1#2{\begingroup\def\par{\endgraf\endgroup\lineskiplimit=0pt}
               \setbox2=\hbox{\large\sc #2}
               \newdimen\tmpht \tmpht \ht2 \advance\tmpht by \baselineskip
               \font\hhuge=Times-Bold at \tmpht
               \setbox1=\hbox{{\hhuge #1}}
               \count7=\tmpht \count8=\ht1
               \divide\count8 by 1000 \divide\count7 by \count8
               \tmpht=.001\tmpht\multiply\tmpht by \count7
               \font\hhuge=Times-Bold at \tmpht
               \setbox1=\hbox{{\hhuge #1}}
               \noindent
                \hangindent1.05\wd1
               \hangafter=-2 {\hskip-\hangindent
               \lower1\ht1\hbox{\raise1.0\ht2\copy1}%
                \kern-0\wd1}\copy2\lineskiplimit=-1000pt}

\def\Kout{\setbox1=\hbox{\Huge\bf K}\hbox to
1.05\wd1{\hspace{.05\wd1}
}}
\def\Sout{\setbox1=\hbox{\Huge\bf S}\hbox to 1.05\wd1{\hspace{.05\wd1}
}}

\title{Energy-Latency Tradeoff for  In-Network Function \\ Computation  in Random  Networks}

\author{Paul Balister$^*$, B\'ela Bollob\'as\footnote{P. Balister and B. Bollob\'as are with the Dept. of Math., Univ. of Memphis, Memphis, TN, USA
Email: {\tt \{pbalistr@,bollobas@msci.\}memphis.edu}. They   are supported   in part by NSF grants DMS-0906634, CNS-0721983 and CCF-0728928, and ARO grant W911NF-06-1-0076. }, Animashree Anandkumar\footnote{A. Anandkumar is with the Center for Pervasive Communications and Computing, Electrical Engineering and Computer Science Dept., University of California, Irvine, USA 92697. Email: {\tt a.anandkumar@uci.edu}. She is  supported by the setup funds at UCI.}, Alan Willsky\footnote{A.S. Willsky is with the Dept. of EECS, Massachusetts Institute of Technology, Cambridge, MA, USA. Email: {\tt willsky@mit.edu}. He is  supported in part by a MURI funded through ARO Grant W911NF-06-1-0076.}}

\begin{document}
\maketitle

\begin{abstract}The problem of designing  policies for in-network function computation with minimum energy consumption subject to a latency constraint   is considered. The scaling behavior of the energy consumption under the latency constraint is analyzed for random networks, where the nodes are uniformly placed in growing regions and the number of nodes goes to infinity.  The special case of sum function computation and its delivery to a designated root node is considered first. A policy which achieves order-optimal average energy consumption in random networks subject to the given latency constraint is proposed. The scaling behavior of the optimal energy consumption depends on the path-loss exponent of wireless transmissions and the dimension of the Euclidean region where the nodes are placed. The policy is then extended to computation of a general class of functions which decompose  according to maximal cliques of a proximity graph such as the $k$-nearest neighbor graph or the geometric random graph.  The modified policy achieves order-optimal energy consumption albeit for a  limited range of latency constraints.
\end{abstract}

\noindent{\bf Keywords: }Function computation, latency-energy tradeoff, Euclidean random graphs, minimum broadcast problem.

\section{Introduction}

A host of emerging networks are pushing the boundaries of scale and complexity. Data centers are being designed to distribute  computation over thousands of machines. Sensor networks are being deployed in larger sizes for a variety of environmental monitoring tasks. These emerging networks  face numerous challenges and the threat of a ``data deluge'' is an important one. The data collected by these networks typically scale rapidly as their size    grows. Routing all the raw data generated in these large networks   is thus not feasible and has poor scaling of  resource requirements.

In this paper, we consider the scenario where  only a   function of the collected raw data  is required at some specific node in the network. Many network applications fall into this category. For instance, in a {\em statistical inference} application, where a decision has to be made based on the collected data, the {\em likelihood} function suffices to make the optimal decision \cite{Anandkumar&Yukich&Tong&Swami:09JSAC}. Such functions can have significantly lower dimensions than the raw data and   can thus considerably reduce the resource requirements for routing.

In this paper, we analyze the scaling behavior of energy and latency for routing and computation of functions in random networks, where the nodes are placed uniformly in growing regions and the number of nodes goes to infinity.   In particular, we address the following questions: how can we exploit the structure of the function to reduce energy consumption and latency? What class of functions can be computed efficiently with favorable scaling of energy and latency requirements? How do the network properties such as the signal propagation model affect the scaling behavior?  What is the complexity for finding the optimal policy with minimum energy consumption under a given latency constraint for function computation?  Are  there simple and efficient policies which achieve order optimal energy consumption? The answers to these questions provide important insights towards engineering in-network computation in large networks.

\subsection{Summary of Contributions}

The contributions of this paper are three-fold. First, we propose policies with efficient energy consumption which compute any function belonging to a certain structured class subject to a feasible latency constraint. Second, we prove order-optimality of the proposed policies   in random networks.
Third, we derive scaling laws for energy consumption in different regimes of latency constraints for different network models. To the best of our knowledge, this is the first work to analyze energy-latency tradeoff for function computation   in large networks. These results provide insight into the nature of functions which are favorable for in-network computation.

We analyze the scaling laws for energy and latency in random networks, where $n$ nodes are placed uniformly in a region of volume (or area) $n$ in $\Rd$, and we let $n \to \infty$. We consider (single-shot) function computation and its delivery to a designated root node. We first consider the class of sum functions, which can be computed via an aggregation tree. We characterize the structural properties of the minimum latency tree and propose an algorithm to build an energy-efficient minimum latency  tree based on successive bisection of the region of node placement. However, minimum latency comes at the expense of energy consumption and we relax the minimum latency constraint. Our modified algorithm achieves order-optimal   energy consumption for any given latency constraint. It is based on the intuition that long-range communication links lower latency but increase energy consumption and the key is to strike a balance between having long-range and short-range communications to achieve the optimal tradeoff.

We then consider the more general class of functions that decompose as a sum of functions over the maximal cliques of a proximity graph, such as the $k$-nearest neighbor graph or the random geometric graph. These functions are relevant in the context of statistical inference of correlated measurements which are drawn from a {\em Markov random field}. See \cite{Anandkumar&Yukich&Tong&Swami:09JSAC} for details. We extend the proposed sum-function policy to this case and prove that it achieves order-optimal energy consumption (up to logarithmic factors) albeit under a limited range of latency constraints. In this range of feasible latency constraints, the energy consumption    is of the same order as sum function computation. Hence, functions based on locally-defined proximity graphs can be computed efficiently with optimal scaling of energy and latency requirements.

We analyze the scaling behavior of energy consumption under different regimes of latency constraints and for different signal propagation models. We assume that the energy consumed scales as $R^\nu$ where $R$ is the inter-node distance and $\nu$ is the path-loss exponent and consider nodes placed in a region in $\Rd$. We prove that in the regime $1\leq\nu<d$, order-optimal energy consumption and minimum latency can both be achieved simultaneously. On the other hand, in the regime $\nu>d$, there is a tradeoff between energy consumption and the resulting latency of computation, and our policy achieves order-optimal tradeoff.


\subsection{Prior and Related Work}

There is extensive literature on in-network processing. Some of the earliest arguments for in-network processing for scalability are presented in \cite{Estrin&etal:99MOBICOM,Pottie&Kaiser:00COMACM}.   The work of Giridhar and Kumar \cite{Giridhar&Kumar:05JSAC} provides a theoretical framework for in-network computation of certain functions such as sum function and  analyze scaling of  capacity  as the network size grows. However, the work in   \cite{Giridhar&Kumar:05JSAC} is concerned with  the rate of information flow   when the function is computed an infinite number of times, while we consider latency of single-shot function computation. Single-shot computation is relevant in applications involving one-time decision making based on a set of measurements. Moreover, we consider a richer class of functions which decompose according to some proximity graph. These are relevant in statistical inference applications with correlated measurements.

For the special case of sum-function computation, the minimum latency is the same as that for the  minimum broadcast problem, where the root has information that needs to be disseminated to all the nodes. Most of the previous work on minimum broadcast problem, e.g., \cite{Ravi:94FOCS,Brosh&etal:07TON}, have focused on obtaining good approximations for minimum latency in arbitrary networks, but do  not address the issue of scaling behavior of latency-energy tradeoff in   random networks. These works also assume that only short-range communication may be feasible for communication. On the other hand, we allow for a few long-range links but focus on obtaining favorable scaling of overall energy consumption.  Works considering latency-energy tradeoff in multihop networks are fewer. For instance, the works in \cite{Lindsey&etal:01,Zorzi&Rao:03MCom,Yu&etal:04INFOCOM} consider energy-latency tradeoff for data collection but without the possibility of in-network computation, which  can  be significantly more expensive. The work in  \cite{Moscibroda:06INFOCOM} considers latency-energy tradeoff but during the deployment phase of the network.

With respect to analysis of energy scaling laws in randomly placed networks, the work in \cite{Zhao&Tong:05JSAC} derives scaling laws for multihop routing without in-network computation. In \cite{Anandkumar&etal:08INFOCOM}, the minimum energy policy for graph-based function computation is first analyzed in the context of statistical inference of correlated measurements and is shown to be NP-hard. An efficient policy is derived based on the Steiner-tree approximation. In \cite{Anandkumar&Yukich&Tong&Swami:09JSAC}, scaling laws for energy consumption are derived for computation of graph-based functions in random networks.
When the function decomposes according to the cliques of   a proximity graph, such as the $k$-nearest neighbor graph or the random geometric graph, it is shown that the function can be computed with\footnote{For any two functions $f(n),g(n)$, $f(n) = O(g(n))$ if there exists a constant $c$ such that $f(n) \leq c g(n)$ for all $n \geq n_0$ for a fixed $n_0\in \Nbb$. Similarly, $f(n) = \Omega(g(n))$ if there exists a constant $c'$ such that $f(n) \geq c'g(n)$  for all $n \geq n_0$ for a fixed $n_0\in \Nbb$, and $f(n) = \Theta(g(n))$ if $f(n)= \Omega(g(n))$ and $f(n) = O(g(n))$.} $\Theta(n)$ energy consumption in random networks, where $n$ is the number of nodes. A simple two-stage computation policy achieves this scaling and is shown to have asymptotically a constant approximation ratio, compared to the minimum energy policy. In this paper, we extend the work to incorporate latency constraints and design policies which minimize energy consumption under the constraints.

\section{System Model}


\subsection{Communication and Propagation Model}\label{sec:prop}

In a wireless sensor network, there are communication and energy constraints. We assume that any node cannot transmit and receive at the same time  (half duplex nodes). We assume that a node cannot receive from more than one transmitter at the same time and similarly, a node cannot transmit simultaneously to more than one receiver. We assume that no other interference constraints are present. This is valid if nearby nodes transmit in orthogonal channels or when they have   idealized directional antenna  which can focus the transmissions within a narrow region around the receiver (e.g., \cite{Yi&etal:03Mobihoc,Shen&etal:TranMC}).   We also assume that nodes are capable of adjusting their transmission power depending on the location of the receiver leading to better energy efficiency.



We assume  unit propagation delays along all the communication links  and negligible processing delays  due to in-network computation  at nodes. For a transmission along edge $(i,j)$ (from  node $i$ to node $j$), the  energy consumption\footnote{Since nodes
only communicate a finite number of bits, we use energy instead of power as the cost measure.} is equal to  $R_{i,j}^\nu$, where $R_{i,j}$ is the Euclidean distance and   typically $\nu \in [2,6]$ for wireless transmissions. In this paper, we allow for any $\nu \geq 1$.

\subsection{Stochastic model of sensor locations}\label{sec:location}
Let $\Bmsc_n\subset \Rd$ denote the $d$-dimensional hypercube $[0,n^{1/d}]^d$ of volume
$n$, and typically $d=2$ or $3$ for sensors placed in an Euclidean region.
 We assume that $n$ sensor nodes (including the root) are
placed uniformly in $\Bmsc_{ n }$ with sensor $i$ located at $V_i \in \Rd$. We denote the set of locations of the $n$ sensors
by $\bfV_n\defeq\{V_1,\ldots,V_n\}$.  For our scaling law analysis, we  let the number of
sensors $n\rightarrow \infty$. Denote the  root node by $ r$, where the computed function needs to be delivered,  and its location by $V_r$.


\subsection{Function Computation Model}
Each sensor node $i$  collects a  measurement   $Y_{i}\in \Yc$, where $\Yc$ is a finite set, and let $\bfY_{n}=\{Y_1,\ldots, Y_n\}$ be the set of  measurements  of $n$ nodes. We assume that the goal of data aggregation  is to ensure that a certain deterministic function\footnote{In general, the function can depend on the locations where the measurements are collected.} $\Psi:(\bfY_{n},\bfV_n)\mapsto \R$ is computable at the root $r$ at the end of the aggregation process. The set of valid aggregation policies $\pi$ is thus given by
\beq \label{eqn:valid}
\mathfrak{F}(\bfV_n;\Psi)
\defeq \{\pi : \mbox{$\Psi(\bfY_{n},\bfV_n)$ computable at  $r$}\}. \eeq
Using the propagation model discussed in Section~\ref{sec:prop}, the total energy consumption of the aggregation process under a policy $\pi\in \mathfrak{F}(\bfV_n;\Psi)$ is
\beq  \Ec^\pi(\bfV_n) := \sum_{e \in G^\pi_n} R_e^\nu,\label{eqn:total_energy}\eeq where $G^\pi_n$ is the set of links used for inter-node communication by the policy. The latency\footnote{We consider one-shot function computation.} of function computation is \beq\label{eqn:latency} L^\pi(\bfV_n;\Psi)
\defeq \inf[t: \mbox{$ \Psi(\bfY_{n},\bfV_n)$ computable at $V_1$ at time $t$}],\eeq where the aggregation process starts at $t=0$. Let  $L^*(\bfV_n;\Psi)$ be the minimum latency   over the set of valid policies.

\begin{figure}[t]\bc\bp\psfrag{r}[r]{\scriptsize Root $r$}\includegraphics[width=1.7in]{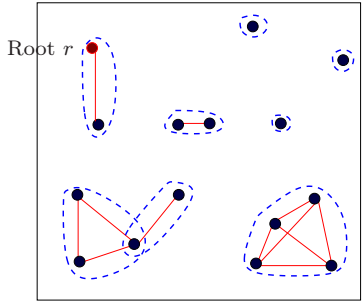}\ep\ec\caption{Example of a function dependency graph $\Gmsc$ and the function decomposes in terms of the maximal cliques of the graph, as represented by dotted lines.}\label{fig:func_clique}\end{figure}

If no further assumptions are made on the function $\Psi$, then   all the measurements $\bfY_{n}$ need to be delivered to the root without any in-network computation. This is expensive both in terms of latency and energy consumption. Typically, the function $\Psi$ decomposes into sub-functions involving only subset of measurements. In this case, in-network computation can be carried out to enable efficient tradeoff between energy consumption and latency of computation. We assume that the function $\Psi$  has the form,
\beq\label{eqn:clique_func_old} \Psi(\bfV_n,\bfY_n) =\sum_{c \in
\Cmsc} \psi_{c}( (Y_i)_{i\in c}),\eeq where $\Cmsc$ is the set of {\em maximal cliques}\footnote{A clique is a complete subgraph and is maximal if it is not contained in a bigger clique.} on some graph $\Gmsc_\Psi$. See Fig.\ref{fig:func_clique} for an example. Note that this graph $\Gmsc_\Psi$ is related to the function $\Psi$ and not with communication links. We refer to $\Gmsc_\Psi$ as the {\em function dependency graph}.

We consider the case when the graph\footnote{In fact, our results hold for a general class of graphs satisfying a certain stabilization property. See \cite{Penrose&Yukich:01AAP} for details and examples.} $\Gmsc$ is either a $k$-nearest neighbor graph ($k$-NNG) or the $\rho$-random geometric graph ($\rho$-RGG) with threshold radius $\rho$, where $k,\rho$ are some fixed constants, independent of the number of nodes $n$. These graphs are relevant choices since many functions are based on proximity of the nodes. For instance, in the context of statistical inference, this corresponds to node measurements being locally dependent according to a {\em Markov random field} with  the given graph $\Gmsc(\bfV_n)$. See \cite{Anandkumar&Yukich&Tong&Swami:09JSAC} for details.


\subsection{Energy-Latency Tradeoff}

Denote the minimum latency for function computation over the set of valid policies by $L^*$, i.e.,
\beq L^*(\bfV_n;\Gmsc_\Psi):=\min_{\pi\in \mathfrak{F}} L^\pi(\bfV_n;\Gmsc_\Psi).\label{eqn:minlatency_def}\eeq
The policy achieving minimum latency $L^*$ can have large energy consumption and similarly, policies with low energy consumption can result in large latency. Hence, it is desirable to have policies that can tradeoff between   energy consumption and the latency of function computation.  We consider  finding a policy with minimum energy consumption subject to a latency constraint,
\beq \Ec^*(\bfV_n;\delta,\Gmsc_\Psi) :=\min_{\pi\in \mathfrak{F}} \Ec^\pi(\bfV_n;\Gmsc_\Psi), \quad \st\,\, L^\pi \leq L^* + \delta,\label{eqn:tradeoff}\eeq where $\delta$ (which can be a function of $n$) is the additional latency suffered in order to reduce energy consumption.
In general, finding \eqref{eqn:tradeoff} is NP-hard for nodes placed at arbitrary locations (since the special case of this problem of finding minimum energy policy with no latency constraints is NP-hard \cite{Anandkumar08INFOCOM}). We instead propose a policy which has energy consumption of the same order as the optimal policy   for randomly placed nodes $\bfV_n$, as $n \to \infty$, and for any given latency constraint.

\section{Sum Function Computation}

A sub-class of functions in \eqref{eqn:clique_func} is the set of sum functions \beq \label{eqn:sum_func}  \Psi(\bfV_n,\bfY_n) =\sum_{i =1
}^n \psi_{i}(  Y_i ),\eeq which have the maximum extent of decomposition over the set of nodes.  Computing sum functions is required in various network applications, e.g., to find the  average value, in distributed statistical inference with statistically independent   measurements \cite{Anandkumar&Yukich&Tong&Swami:09JSAC}, and so on.


\subsection{Preliminaries}

We first discuss the policy to achieve minimum latency $L^*(\bfV_n;\Psi)$ in \eqref{eqn:minlatency_def} for sum function computation without considering the energy consumption. In this   case, the minimum latency  does not depend on the position of the nodes $\bfV_n$ but  only on the order of scheduling of the various nodes, i.e., $L^*(\bfV_n;\Psi) = L^*(n)$. Moreover, the minimum latency $L^*(n)$ can be achieved via data aggregation along a spanning tree $T^*(n)$, directed towards root $r$.

\begin{figure}[t]\bc\bp \psfrag{r}[l]{\scriptsize Root $r$}\psfrag{1}[l]{\scriptsize$1$}\psfrag{2}[l]{\scriptsize$2$}\psfrag{k}[l]{\scriptsize$k$}\psfrag{T1}[l]{\scriptsize $T_1$}\psfrag{...}[c]{...}
\psfrag{T2}[l]{\scriptsize $T_2$}\psfrag{Tk}[l]{\scriptsize $T_k$}
 \includegraphics[width=1.7in]{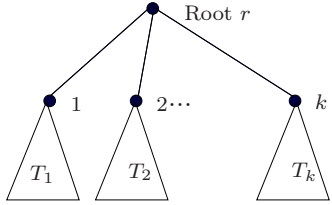}\ep\ec
\caption{The latency for an aggregation tree can be obtained iteratively by considering subtrees. See Proposition~\ref{prop:subtree}.}\label{fig:subtree}
\end{figure}

\begin{figure}[t]\bc\bp \psfrag{r}[l]{\scriptsize Root $r$}\psfrag{1}[l]{\tiny$1$}\psfrag{2}[l]{\tiny$2$}\psfrag{3}[l]{\tiny$3$}\psfrag{4}[l]{\tiny$4$}
 \includegraphics[width=1.7in,height=1.1in]{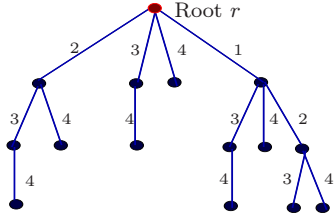}\ep\ec
\caption{The min. latency tree $T^* $  with edge level labels. See Alg.\ref{alg:minlatency}.}\label{fig:mintree}
\end{figure}

For data aggregation along any directed spanning tree $T$, each node waits to receive data from its children (via incoming links),  computes the sum of the values (along with its own measurement) and then forwards the resulting value along the outgoing link. See Fig.\ref{fig:mintree} for an example. Let $L_T$ be the resulting latency along tree $T$. We now make a simple observation. See also Fig.\ref{fig:subtree}.

\begin{proposition}[Latency along a tree]\label{prop:subtree}For a spanning tree $T$ with root $r$, the latency $L_T$ is given by
 \beq L_T = \max_{i=1,\ldots, k} \{i + L_{T_i}\},\label{e:1}\eeq where $T_i$ is the subtree rooted at node $i$, and $1,\ldots, k$ are the children $\Cc(r;T)$ of the root node $r$ ordered such that $L_{T_1}\geq L_{T_2}\ldots\geq L_{T_k}$.
\end{proposition}

\bprf Indeed, after time $L_{T_i}$, information from $1,\dots,i$ has still not
been sent to the root, and this will take at least time $i$,
so $L_T\geq  i + L_{T_i} $ for all $i=1,\ldots,k$. Conversely, there is a simple
policy with latency in \eqref{e:1} which aggregates along the subtrees $T_i$ with latency $L_{T_i}$ and then node $i$ sends its data to the root $r$ at time slot $L_T-i$.  \eprf

Using \eqref{e:1} we can thus effectively compute the latency of any given rooted tree $T$.  We now provide the result on the minimum latency $L^*(n)$ and the construction of the tree $T^*(n)$ achieving it. This has been previously analyzed in the context of minimum broadcast problem  \cite{Ravi:94FOCS}, where the root has information that needs to be disseminated to all the nodes.

\bl[Minimum Latency Tree]\label{lemma:minlatency}The minimum latency for sum function computation over $n$ nodes is $L^*(n) =  \lceil\log_2 n\rceil$. Equivalently, the maximum number of
 vertices in a tree with latency at most\/ $L$ is $2^L$.\el

\bprf See Appendix~\ref{proof:minlatency}.\eprf

There is a unique minimum latency tree\footnote{Note that the balanced binary tree on $n$ nodes has latency $2\lceil\log_2 (n+1)\rceil-2$, which is about twice $L^*(n)$.}  $T^*(n)$ up to a permutation on the nodes. The minimum latency tree can be constructed recursively as explained in Algorithm~\ref{alg:minlatency}. The algorithm runs for $L^*(n)$ steps and in each step, a child is added to each node already in the tree.  An example of the minimum latency tree is shown in Fig.\ref{fig:mintree}.

\begin{algorithm}[tb!]
\caption{Min. latency tree $T^*(n)$.}
\label{alg:minlatency}
\begin{algorithmic}[1]
\REQUIRE nodes $N=\{1,\ldots, n\}$, root node $r$. $\Cc(i;T)$ denotes children of node $i$. $\Sc(k;T)$ denotes level $k$ edges in $T$. For any set $A$, let $A\overset{\cup}{\leftarrow} \{r\}$ denote $A \leftarrow A\cup \{r\}$.
\ENSURE  $T^*(n)$.
\STATE Initialize  set $A=\{r\}$ and $T^*=\{r\}$.
\FOR{$k=1,\ldots,\lceil \log_2 n\rceil$}\STATE $B\leftarrow A$.
\FORALL{$i\in B$}
\IF{$N\setminus A \neq \emptyset$}
\STATE For some $j \in N\setminus A$, $\Cc(i;T^*)\overset{\cup}{\leftarrow}j$ ($j$ is now a child of $i$), $\Sc(k;T^*) \overset{\cup}{\leftarrow}(i,j)$ (level $k$ edges) and  $A\overset{\cup}{\leftarrow} j $.
\ENDIF
\ENDFOR
\ENDFOR
\end{algorithmic}
\end{algorithm}

\begin{figure}[t]\bc\bp \psfrag{r}[r]{\scriptsize Root $r$}\psfrag{1}[l]{\tiny$1$}\psfrag{2}[l]{\tiny$2$}\psfrag{3}[l]{\tiny$3$}\psfrag{4}[l]{\tiny$4$}
 \includegraphics[width=1.7in]{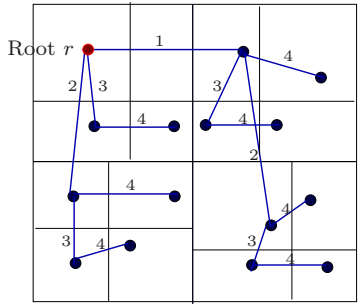}\ep\ec
\caption{The min. latency tree $T^*(\bfV_n) $ over $ 15$ nodes  with edge level labels placed in square region. See Alg.\ref{alg:minlatency_energy}.}\label{fig:euctree}
\end{figure}

\subsection{Policies for Energy Latency Tradeoff}

We now propose a policy for sum function computation with order-optimal energy consumption subject to a given latency constraint  in \eqref{eqn:tradeoff}. Note that the minimum latency tree $T^*(n)$ does not depend on the node locations and any permutation of the nodes on the tree (with the root fixed) results in the same latency. On the other hand, the energy consumption depends on the node locations. We propose an energy-efficient minimum-latency  tree $T^*(\bfV_n)$ in Algorithm~\ref{alg:minlatency_energy}, depending on the node locations. This will be  later proven to achieve order-optimal energy consumption for uniformly placed nodes. We first note some definitions used in the sequel.

\noindent{\em Definitions: }For a rooted tree $T$ and a node $i$, let $\Cc(i;T)$ denote the children of $i$. The {\em level} $l(e;T)$ of a link $e$ in an aggregation tree $T$ is given by $L_T - t_e$, where $t_e$ is the time at which  data is transmitted along link $e$ (time $0$ is the start of the aggregation process). Note that the level depends on both the tree structure and the transmission schedule on the tree. Let \beq \Sc(k;T) := \{e:  l(e;T)=k, e \in T\}.\label{eqn:levelk}\eeq be the set of level $k$ edges in tree $T$. See Fig.\ref{fig:mintree}.
Let $\SP_{l}(i,j;\bfV_n)$  denote the least-energy path\footnote{Note that the least-energy path depends on the path-loss exponent $\nu$ and for larger $\nu$, multi-hop routing is more energy efficient than direct transmissions.} between $i$ and $j$  with at most $l\geq 0$ intermediate nodes when the node locations are $\bfV_n$. For a rectangular region $Q\subset \Rd$ containing a subset of nodes and a reference node $i$ such that $V_i\in Q$, let $\Bc_1(Q;i), \Bc_2(Q;i)$ be the two halves when bisected along the coordinate with the largest extent such that $\Bc_1(Q;i)$ and $\Bc_2(Q;i)$ have equal number of nodes with $V_i\in \Bc_1(Q;i)$.

We propose an energy-efficient minimum-latency  tree $T^*(\bfV_n)$ in Algorithm~\ref{alg:minlatency_energy}, and, prove in Section~\ref{sec:guarantees} that $T^*(\bfV_n)$ achieves order-optimal energy consumption  for uniformly placed nodes subject to the minimum latency constraint. In Algorithm~\ref{alg:minlatency_energy}, every added node in the tree picks a new child, as in Algorithm~\ref{alg:minlatency}, but now the children are chosen based on the node locations. Specifically, in the first iteration, the region of node placement is bisected (with equal number of nodes in each half), and the root chooses a child in the other half. The region assigned to the root is now the half-region (where it is located), while the added child node is assigned the other half-region. The subsequent iterations proceed in a similar manner and each node bisects its assigned region into two halves and picks a child in the other half, and updates the assigned regions.


\begin{algorithm}[tb!]
\caption{Min. lat. tree $T^*(\bfV_n)$ with order opt. energy}
\label{alg:minlatency_energy}
\begin{algorithmic}[1]
\REQUIRE Locations of nodes: $\bfV_n=\{V_1,\ldots, V_n\}$, root node $r$, $\Bmsc_n\subset \Rd$: region where the nodes  are placed.   $\Cc(i;T)$ denotes children of node $i$. $\Sc(k;T)$ denotes level $k$ edges in $T$. For a rectangular region $Q$ and a node $i$ with $V_i\in Q$, let $\Bc_1(Q;i)$ and $\Bc_2(Q;i)$ be the two halves with $V_i\in \Bc_1(Q;v)$.  For any set $A$, let $A\overset{\cup}{\leftarrow} \{r\}$ denote $A \leftarrow A\cup \{r\}$.
\ENSURE  $T^*(\bfV_n)$.
\STATE Initialize $A\leftarrow \{r\}$. $R_r\leftarrow Q_n$.
\FOR{$k=1,\ldots,\lceil \log_2 n\rceil$}
\STATE $B\leftarrow A$.
\FORALL{$i \in B$}
\IF{$\bfV_n\cap \Bc_2(R_i;i) \neq \emptyset$}
\STATE For some node $j$ s.t. $V_j \in \Bc_2(R_i;i)$, $\Cc(i;T^*)\overset{\cup}{\leftarrow}j$, $\Sc(k;T^*) \overset{\cup}{\leftarrow} (i,j)$, $A\overset{\cup}{\leftarrow}\{j\}$, $R_j \leftarrow \Bc_2(R_i;i)$ and $R_i \leftarrow \Bc_1(R_i;i)$.
\ENDIF
\ENDFOR
\ENDFOR
\end{algorithmic}
\end{algorithm}

The algorithm~\ref{alg:minlatency_energy} considered energy-efficient policy under the minimum latency constraint. We now present the policy $\pi^\Sum$  for any given latency constraint in Algorithm~\ref{alg:tradeoff}.   The difference between the two cases is that in the latter case, a lower energy consumption is achieved by exploiting the relaxed latency constraint. Intuitively, long-range (direct) communication entails more energy consumption than multi-hop routing, especially when the path-loss exponent $\nu$ is large.  On the other hand, latency is increased due to multihop routing. The key is to carefully convert some of the long-range links in $T^*(\bfV_n)$ into multi-hop routes to lower the energy consumption and take advantage of the additional allowed latency.

In Algorithm~\ref{alg:tradeoff}, the regions are bisected and new nodes are chosen as children, as in Algorithm~\ref{alg:minlatency_energy}. But instead of directly linking the nodes in the two hops, the least-energy route is chosen with at most $w_k$ intermediate routes, where $w_k$ is a fixed weight. The nodes that are already added in this manner are not considered for addition as children in the subsequent iterations. In general, the resulting set of communication links is not a tree, since the least-energy paths constructed in different iterations may share the same set of nodes. But the sum function computation can be carried out on similar lines, as on an aggregation tree. We now relate the weights $(w_k)$ with the latency of the resulting policy $\pi^\Sum$ in Algorithm~\ref{alg:tradeoff}.


\begin{proposition}[Latency under Algorithm~\ref{alg:tradeoff}]\label{lemma:latency}The aggregation policy $\pi^\Sum$ in  Algorithm~\ref{alg:tradeoff} for a given set of weights $\bfw$ achieves a latency of \[ L^{\pi^\Sum}(n) \leq L^*(n) + \sum_{k=0}^{\lceil \log_2 n \rceil-1} w_k.\]\end{proposition}

\bprf There are at most $\lceil \log_2 n \rceil$ iterations and the total delay is
\[\sum_{k=0}^{\lceil \log_2 n \rceil-1} (1+w_k) =  L^*(n) + \sum_{k=0}^{\lceil \log_2 n \rceil-1} w_k.\] \eprf

Thus, the  weights $(w_k)$ can be chosen to satisfy any given  latency constraint and we have a policy $\pi^\Sum$ for sum function computation given any feasible latency constraint. The analysis of energy consumption under   $\pi^\Sum$ for a given set of weights is not straightforward to analyze and forms the main result of this paper. This is discussed in the next section.


\begin{algorithm}[tb!]
\caption{Latency-energy tradeoff policy $\pi^\Sum(\bfV_n; \bfw)$. }
\label{alg:tradeoff}
\begin{algorithmic}[1]
\REQUIRE Locations of nodes: $\bfV_n=\{V_1,\ldots, V_n\}$, root node $r$,   and set of weights $w_k$ for $k=0,\ldots, \lceil \log_2 n \rceil-1$. For a rectangular region $Q$ and node $v\in Q$, let $\Bc_1(Q;v)$ and $\Bc_2(Q;v)$ be the two halves with $v\in \Bc_1(Q;v)$.  Let $\SP_{l}(i,j;\bfV_n)$ be $l$-hop least-energy path. $\Bmsc_n\subset \Rd$: region where the nodes  are placed. For any set $A$, let $A\overset{\cup}{\leftarrow} \{r\}$ denote $A \leftarrow A\cup \{r\}$.
\ENSURE  $G^{\pi^\Sum}$:  communication links used by policy $\pi^\Sum$.
\STATE Initialize  $A_1,A_2\leftarrow \{r\}$. $R_r\leftarrow Q_n$.
\FOR{$k=0,\ldots,\lceil \log_2 n\rceil-1$}
\STATE $B\leftarrow A_1$
\FORALL{$i \in B$}
\IF{$(\bfV_n\cap \Bc_2(R_i;i))\setminus  A_2  \neq \emptyset$}
\STATE Pick    $j$ s.t. $V_j \in \Bc_2(R_i;i)\setminus  A_2 $, $A_1 \overset{\cup}{\leftarrow}\{j\}$,\\ $G^{\pi^\Sum}\overset{\cup}{\leftarrow}\SP_{w_k}(i,j;\bfV_n)$,  $A_2\overset{\cup}{\leftarrow} \SP_{w_k}(i,j;\bfV_n)$, \\ $R_j \leftarrow \Bc_2(R_i;i)$ and $R_i \leftarrow \Bc_1(R_i;i)$.
\ENDIF
\ENDFOR
\ENDFOR
\end{algorithmic}
\end{algorithm}

\subsection{Order-Optimality Guarantees}\label{sec:guarantees}

To achieve optimal energy-latency tradeoff according to \eqref{eqn:tradeoff}, we choose weights $w_k$ in Algorithm~\ref{alg:tradeoff}, for $k=0,\ldots, \lceil \log_2 n\rceil-1$, as \bcase{w_k =} \lfloor \zeta \delta 2^{k(1/\nu - 1/d)}\rfloor & if $\nu> d$, \nn \\ 0 &o.w.\label{eqn:weights}\ecase where $\delta$ is the additional latency allowed in \eqref{eqn:tradeoff}, $\nu$ is the path-loss factor for energy consumption in \eqref{eqn:total_energy} and $d$ is the dimension of Euclidean space where the nodes are placed. The normalizing  constant $\zeta$ is chosen as \bcase{\zeta=} 1- 2^{1/\nu- 1/d}, & if $\nu\geq d$, \nn\\ \lceil \log_2 n \rceil^{-1}, &$\nu=d$ ,\ecase
so that $\sum_{k=0}^{\lceil \log_2 n \rceil-1} w_k \leq \delta$. Hence, from Lemma~\ref{lemma:latency}, the weights in \eqref{eqn:weights} result in a policy $\pi^\Sum$ with latency $L^*(n) + \delta$. We now provide the scaling behavior of optimal energy consumption as well the order-optimality result for $\pi^\Sum$.

\begin{theorem}[Energy-Latency Tradeoff]\label{t:main}
For a given additional latency constraint $\delta=\delta(n)\ge0$ and fixed path-loss factor $\nu> 1$ and dimension $d\geq 1$, as the number of nodes  $n\to\infty$, the minimum energy consumption for sum function computation satisfies
\[ \E(\Ec^*(\bfV_n;\delta))\!\!=\!\!
  \begin{cases}\!\!
  \Theta(n) & \!\! \nu<d,\\ \!\!
  O\big(\max\{n,n(\log n)(1+\frac{\delta}{\log n})^{1-\nu}\}\big) &\!\!\nu=d ,\\ \!\!
  \Theta\big(\max\{n,n^{\nu/d}(1+\delta)^{1-\nu}\}\big) & \!\!\nu>d,
  \end{cases}
 \] where the expectation is over the locations  $\bfV_n$ of $n$ nodes
 chosen uniformly at random in $[0,n^{1/d}]^d$ and is achieved by the policy $\pi^\Sum$ in Algorithm~\ref{alg:tradeoff}    for weights given by \eqref{eqn:weights}.
\end{theorem}

\begin{figure}[t]\bc\bp\psfrag{n}[l]{\scriptsize$n$}\psfrag{E}[l]{\scriptsize$\Ec^*$}
\psfrag{nu1}[l]{$\nu<d$}\psfrag{nu2}[r]{$\nu>d$}
\includegraphics[width=1.5in]{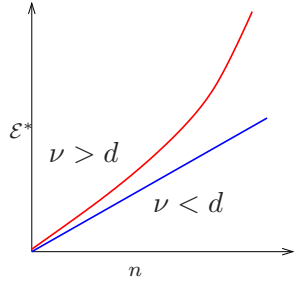}\ep\ec\caption{Scaling of minimum total energy $\Ec^*$  in different regimes of path loss $\nu$ and dimension $d$. See Theorem~\ref{t:main}.}\label{fig:energy_graph} \end{figure}

\noindent{\em Remarks:}

\noindent{\em(i)}The policy $\pi^\Sum$ in Algorithm~\ref{alg:tradeoff}   thus achieves order-optimal energy consumption under any feasible latency constraint when $\nu\neq d$. For the case $\nu=d$, we show that $\Ebb[\Ec^*_n(\bfV_n;\delta)] = \Omega (n)$ while the energy consumption under $\pi^\Sum$ is the upper bound in Theorem~\ref{t:main}, given by $O(n \log n)$, i.e., the energy consumption is at most only logarithmically worse than the lower bound.

\noindent{\em(ii)}The result of Theorem~\ref{t:main} holds even if the latency constraint is relaxed to an average constraint, i.e., \[ \Ebb[L^\pi(\bfV_n)]\leq L^*(n) + \delta.\]

From Theorem~\ref{t:main}, the energy consumption has different behaviors in the regimes $\nu<d$ and $\nu>d$, as represented in Fig.\ref{fig:energy_graph}, and we discuss this below.

\noindent{\em Case $\nu<d$: }In this regime, the path-loss factor is low, and hence, long-range  transmission does not suffer a high penalty over   multihop routing. This is also favorable for latency performance and hence, in this regime minimum latency of $\lceil \log_2 n \rceil$ can be achieved with $\Theta(n)$ energy consumption. Note that the aggregation tree with the minimum energy consumption is the minimum spanning tree (MST) and the expected energy consumption for MST under uniform node placement is also $\Theta(n)$. Hence, our policy $\pi^\Sum$ achieves both order-optimal energy and minimum latency simultaneously in the regime  $\nu<d$.

\noindent{\em Case $\nu>d$: }In this regime, multihop routing is much more favorable over direct transmissions with respect to energy consumption while direct transmissions are  favorable for low latency. Hence,  both low energy and low latency cannot be achieved simultaneously in this regime. Our policy $\pi^\Sum$ achieves order-optimal energy consumption subject to a given latency constraint in this regime. Note that typically, for sensor networks placed in two-dimensional area $(d=2)$ with wireless transmissions $(\nu\in [2,6])$, this regime is of interest.

\noindent{\em Comparison with $\mst$: }If the minimum spanning tree (MST) is used for aggregation, it results in minimum energy consumption which is $\Theta(n)$ under random node placement. However, the expected latency of aggregation along the MST  is at least the depth of the MST and hence, the latency and energy satisfy,  \beq \Ebb[L^{\pi^\tmst}(\bfV_n)] = \Omega(n^{1/d}),\,\, \,\Ebb[\Ec^{\pi^\tmst}(\bfV_n)]=
\Theta(n).\eeq In contrast, under our policy $\pi^\Sum$, we can obtain, when $\nu<d$,\beq \Ebb[L^{\pi^\Sum}(\bfV_n)] =  \lceil\log_2 n\rceil,\,\,\,\Ebb[\Ec^{\pi^\Sum}(\bfV_n)]=
\Theta(n).\eeq For the case when $\nu>d$, our policy achieves
   \beq \Ebb[L^{\pi^\Sum}(\bfV_n)] =  \Theta(n^{\frac{\nu/d-1}{\nu-1}}),\,\,\,\Ebb[\Ec^{\pi^\Sum}(\bfV_n)]=
\Theta(n),\eeq by setting $\delta = n^{\frac{\nu/d-1}{\nu-1}}$ in Theorem~\ref{t:main} Thus, our policy $\pi^\Sum$ is especially advantageous over using the MST when the path loss $\nu$ is small. Moreover, the policy $\pi^\Sum$ can be designed based on the latency requirements while the MST cannot be easily modified to satisfy them.

\section{General Function Computation}

We now extend the latency-energy tradeoff policy to undertake general function computation. Recall that we consider the class of functions of the form\beq\label{eqn:clique_func} \Psi(\bfV_n,\bfY_n) =\sum_{c \in
\Cmsc(\bfV_n)} \psi_{c}( (Y_i)_{i\in c}),\eeq where $\Cmsc(\bfV_n)$ is the set of {\em maximal cliques}  on a graph $\Gmsc(\bfV_n)$, known as the function dependency graph and the functions $\psi_c$ are clique functions. We consider the case when the graph\footnote{In fact, our results hold for a general class of graphs satisfying a certain stabilization property. See \cite{Penrose&Yukich:01AAP} for details and examples.} is either a $k$-nearest neighbor graph ($k$-NNG) or the $\rho$-random geometric graph ($\rho$-RGG) with threshold radius $\rho$, where $k,\rho$ are some fixed constants, independent of the number of nodes $n$.

Note that the function $\Psi$ in \eqref{eqn:clique_func} now depends on the location of the nodes $\bfV_n$ which is not the case with the sum function. Hence, the latency-energy analysis has to take this into account. We   propose modifications to the latency-energy tradeoff policy $\pi^\Sum$ to enable general function computation and then prove its order-optimality for latency-energy tradeoff.

\subsection{Preliminaries}

The extent of decomposition of function $\Psi$ depends on the {\em sparsity} (number of edges) of the function dependency graph $\Gmsc$. We first make a simple observation that the energy consumption and latency  increase with more edges in $\Gmsc$.

\begin{proposition}[Energy consumption and sparsity of $\Gmsc$]\label{prop:sparsity}The minimum energy in \eqref{eqn:tradeoff} required to compute functions of the form in \eqref{eqn:clique_func} under a fixed additional latency constraint $\delta\geq 0$ with dependency graphs $\Gmsc$ and $\Gmsc'$  satisfies\beq \Ec^*(\bfV_n;\delta,\Gmsc) \geq \Ec^*(\bfV_n;\delta,\Gmsc'), \quad\mbox{when }\Gmsc\supset \Gmsc'.\eeq  \end{proposition}

\bprf  Let $\Cmsc$ (resp. $\Cmsc'$ be the set of maximal cliques in $\Gmsc$ (resp. $\Gmsc'$).  Since $\Gmsc'\subset \Gmsc$, each new clique   $c \in \Cmsc\setminus \Cmsc'$ replaces a  smaller set of  cliques     $c'\subset c, c'\in \Cmsc'$.  The latency and energy consumption for any valid policy   in computing the clique function $\psi_{c }((Y_i)_{i \in c })$ is at least that of computing   $\sum\limits_{c'\subset c,c' \in \Cmsc'} \psi_{c'}(\bfY_{c'})$, since this is a further decomposition of $\psi_c$. Hence, the result.\eprf

Hence, the ability to obtain efficient scaling of latency and energy consumption for function computation depends on the sparsity of the function dependency graph $\Gmsc$.   The extreme case of a trivial  graph $(\Gmsc=\emptyset)$ is the sum function, analyzed in the previous section, while the other extreme is the complete graph $(\Gmsc=K_n)$, where there is no decomposition of the function. In the latter case, no in-network computation is possible and all the measurements $\bfY_n$ need to be routed to the root via least-energy paths. We have the following scaling in this scenario.

\begin{proposition}[Scaling Under No Computation]The minimum latency and minimum energy (with no latency constraint) for computation of a function with dependency graph $K_n$ satisfies \[ \Ebb[\Ec^*(\bfV_n;\infty,K_n)]= \Theta(n^{1+ 1/d}),\,\, \Ebb[L^*(\bfV_n;K_n)] = \Omega(n).\] \end{proposition}

The result on minimum energy follows from the scaling behavior of    energy    for least-energy path routing to the root under uniform node placement  \cite{Baccelli:09NOW}. The latency of function computation is at least $n$ since the root can receive at most one measurement value at each timestep and there is no aggregation of the measurements.
Hence, we can expect efficient scaling of energy and latency only in case of computation of functions with sparse dependency graphs $\Gmsc$.

Moreover, the energy consumption also depends on the edge lengths of the function dependency graph $\Gmsc$. Intuitively, when the graph $\Gmsc$ has local edges, the clique functions $\psi_c$ can be computed locally resulting in low energy consumption. This holds for  the proximity graphs such as the $k$-NNG and the $\rho$-RGG ensure under consideration. We propose policies for latency-energy tradeoff which are efficient for such locally-defined dependency graphs.

\subsection{Policy for Latency-Energy Tradeoff}

\begin{algorithm}[tb!]
\caption{Policy $\pi^\Gen(\bfV_n; \bfw,\Cmsc)$ for general functions. }
\label{alg:tradeoff_gen}
\begin{algorithmic}[1]
\REQUIRE Locations of nodes: $\bfV_n=\{V_1,\ldots, V_n\}$, root node $r$,   $\Cmsc$: set of maximal cliques of function dependency graph $\Gmsc(\bfV_n)$. For each $c\in \Cmsc$, $\Pc(c)$ is the processor (node computing clique function $\psi_c$). For any set $A$, let $A\overset{\cup}{\leftarrow} \{r\}$ denote $A \leftarrow A\cup \{r\}$.
\ENSURE  $\pi^\Gen$  policy with data forwarding links $F^{\pi^\Gen}$ and aggregation links  $G^{\pi^\Sum}$.
\FORALL{clique $c \in \Cmsc$}
\STATE For node $i\in c$ with smallest label, $\Pc(c)\leftarrow i$.
\STATE For all nodes $j \in c$, $j\neq i$, $F^{\pi^\Gen}\overset{\cup}{\leftarrow} (j,i)$.
\ENDFOR
\STATE Let $l_e$ be the color for edge $e \in F^{\pi^\Gen}$ under proper edge coloring with colors $l=1,2,\ldots \Delta+1$.
\FOR{$t=0$ to $\Delta$}
\STATE Send measurements using links in $F^{\pi^\Gen}$ of color $t+1$.
\ENDFOR
\STATE Find sum of clique functions  using $\pi^\Sum$ from Algorithm~\ref{alg:tradeoff}.
\end{algorithmic}
\end{algorithm}

We now extend the policy $\pi^\Sum$   in Algorithm~\ref{alg:tradeoff} for general function computation as a two-stage policy $\pi^\Gen$. In the first stage known as the data forwarding stage, the clique functions $\psi_c$ are computed locally within each maximal clique $c\in \Cmsc$ of the graph $\Gmsc$ as follows: a {\em clique processor} is chosen  as a clique member with the smallest label (under arbitrary labeling of nodes) and other clique members communicate their measurements to the processor via direct transmissions. The transmissions are scheduled as follows: the set of forwarding links $F^{\pi^\Gen}$ are assigned colors $l=1,2,\ldots, \Delta +1$ under a proper edge coloring. At times $t=0,1,\ldots, \Delta$, transmissions along links of color $t+1$ are scheduled simultaneously. In the second stage, the aggregation policy $\pi^\Sum$ in Algorithm~\ref{alg:tradeoff}  is used for computing the sum of the clique function values at the processors (and nodes other than processors   do not have their own values for aggregation but participate in the process). This is summarized in Algorithm~\ref{alg:tradeoff_gen}.

We obtain the following result for energy-latency tradeoff\footnote{The latency constraint $L^\pi \leq L^* +\delta$ is required a.s. over the realization of points $\bfV_n$.} for general function computation. Let $\Delta(\Gmsc)$ denote the maximum degree of the function dependency graph $\Gmsc$ in \eqref{eqn:clique_func}, which is either the $k$-NNG or the $\rho$-RGG, where $k$ and $\rho$ are fixed constants.


\begin{theorem}[Energy-Latency Tradeoff]\label{t:clq}
For a given additional latency constraint $\delta\ge \Delta(\Gmsc)+1$ in \eqref{eqn:tradeoff}, the   energy consumption    for function computation of the form \eqref{eqn:clique_func} with dependency graph $\Gmsc$ under the two-stage policy $\pi^\Gen$  satisfies
\[ \E(\Ec^{\pi^\Gen} (\bfV_n;\delta,\Gmsc))= \Theta(\E(\Ec^*(\bfV_n;\delta-( \Delta+1) ,\emptyset))),
 \] where the expectation is over the locations  $\bfV_n$ of $n$ nodes
 chosen uniformly at random in $[0,n^{1/d}]^d$ and the right-hand side is the minimum energy consumption for sum function computation under latency constraint of $\delta- (\Delta+1)$ which is given by Theorem~\ref{t:main}.
\end{theorem}

\noindent{\em Remarks: }

\noindent{\em(i)} The policy $\pi^\Gen$ achieves order-optimal energy consumption for cases $\nu<d, \delta\ge\Delta+1$ and $\nu>d, \delta \gg \Delta$. This is because the minimum energy consumption $\Ec^*(\bfV_n;\delta,\Gmsc)$ is lower bounded by the minimum energy for sum function computation from Proposition~\ref{prop:sparsity}. Theorem~\ref{t:main} provides the scaling for minimum energy for sum function computation. Comparing it with the energy under $\pi^\Gen$ policy in Theorem~\ref{t:clq}, we note that they are both $\Theta(n)$ when $\nu<d$. For the case $\nu>d$, they are still of the same order if the maximum degree $\Delta(\Gmsc)$ is small compared to additional latency constraint $\delta$.

\noindent{\em(ii)} The maximum degrees of $k$-NNG and  $\rho$-RGG   satisfy
\beq \Delta(k\mbox{-NNG}) = (c_d +1)k, \quad \Delta(\rho\mbox{-RGG}) = \Theta(\frac{\log n}{\log \log n}),\label{eqn:maxdegree}\eeq
where $c_d$ is a constant (depending only on $d$). See  \cite[Cor. 3.2.3]{Miller&etal:97JACM} and \cite[Thm. 6.10]{Penrose:book}. Hence, for these graphs,   $\pi^\Gen$ policy is order-optimal (up to logarithmic factors) for any path-loss factor $\nu\neq d$ and under any additional latency constraint $\delta\geq \Delta(\Gmsc)+1$. The above discussion also implies that   the minimum energy for sum function computation and general function computation are of the same order for $k$-NNG and $\rho$-RGG dependency graphs.  Hence, these functions are amenable to efficient latency-energy tradeoff.

\noindent{\em(iii)}The policy $\pi^\Gen$ can achieve a latency of $\lceil \log_2n\rceil + \Delta(\Gmsc)+1$. Finding the policy with minimum  latency $L^*$ in \eqref{eqn:minlatency_def} for general function computation is   NP-hard.  However, we have $L^* \geq \lceil \log_2n\rceil $, since the minimum latency cannot be smaller than that required for sum function computation. We  ensure that an additional latency constraint of $\delta $ is satisfied in \eqref{eqn:tradeoff} by relaxing the constraint as $L \leq \lceil \log_2n\rceil + \delta$. Since $\pi^\Gen$ can only achieve latencies greater than $\lceil \log_2n\rceil + \Delta(\Gmsc)+1$, we can only ensure that constraints $\delta\geq \Delta(\Gmsc)+1$ are met.


\section{Conclusion}

In this paper, we considered energy-latency tradeoff for function computation in random networks. While designing optimal tradeoff policies in arbitrary networks is intractable, we proposed simple and easily implementable policies which have order-optimal performance and are relevant in large networks. We analyzed the scaling behavior of energy consumption under a latency constraint for computation and showed that it depends crucially on the path-loss exponent of signal propagation, the dimension of the Euclidean region where the nodes are placed and the extent to which the function is decomposable. For functions which decompose according to cliques of a proximity graph such as the $k$ nearest-neighbor graph or the random geometric graph, efficient tradeoff can be achieved and the energy and latency having optimal scaling behaviors.

This work opens up an array of important and challenging questions which warrant further investigation. While, we considered exact computation of a deterministic function, we expect that relaxing these assumptions will lead to a significant improvement of energy and latency scaling. We assumed  single-shot data aggregation. Extensions to the setting of continuous monitoring and collection, where block coding is possible is of interest.
We considered a single root node as the destination for the computed function, while in reality different nodes may require different functions to be computed. An extreme case of this scenario is the {\em belief propagation} (BP) algorithm which requires computation of  {\em maximum a posteriori} (MAP) estimate at each node  based on all the  measurements, which are drawn from a Markov random field.  Considering scenarios between these extremes and designing efficient schemes for energy-latency tradeoff is extremely relevant to many network applications.


\begin{appendix}

\section{Proof of Lemma~\ref{lemma:minlatency}}\label{proof:minlatency}

We prove by induction on $L$ that the maximum number of vertices in
a tree of latency at most $L$ is exactly $2^L$. This is clear for $L=0$
as such a tree must consist of just the root. Now assume $L>0$ and suppose
$T$ is a tree with latency~$L$. Consider the edges that transmit information
at the last time step $L$. Clearly these must transmit to the root~$r$. But
the root can only receive information from one child at a time. Thus there is
precisely one edge $(r,i)$ along which information is transmitted at time~$L$.
Removing the edge $(r,i)$ splits the tree $T$ into two trees $T_r$ and $T_{i}$
rooted at $r$ and $i$ respectively. For all the data to be received at $r$
in time $L$, all the data must be received at either $r$ or $i$ by time $L-1$.
Thus both $T_r$ and $T_{i}$ are trees of latency at most $L-1$. By induction
$T_r$ and $T_{i}$ have at most  $2^{L-1}$ vertices. Thus $T$ has at most
$2^{L-1}+2^{L-1}=2^L$ vertices. Conversely, given two copies of a rooted tree
on $2^{L-1}$ vertices with latency $L-1$, one can construct a tree with latency
$L$ on $2^L$ vertices by joining the roots $r$, $i$ of these two trees with an
edge $(r,i)$, and declaring one of the two roots, say $r$, to be the root of the
resulting tree. The transmission protocol to achieve latency $L$ is simply to
follow the protocols on each tree for the first $L-1$ steps, and then transmit
all data at $i$ from $i$ to $r$ at time step~$L$.

As any rooted subtree of a tree $T$ has latency at most $L_T$, it is clear
that the minimum latency of any tree on $n$ vertices is $L=\lceil\log_2 n\rceil$,
and this can be achieved by taking any rooted subtree of the tree on $2^L$ vertices
constructed above.\qed

\section{Proof  of Lower Bound in Theorem~\ref{t:main}}\label{proof:main}

Note that for $\nu< d$, since the MST has energy $\Theta(n)$, the result follows. For the case $\nu>d$, consider an arbitrary spanning tree with root $r$. Consider the path
$P_u$ from $r$ to $u$ in the tree. Let $R(P_u)$ be the length of~$P_u$,
i.e., the sum of the lengths of the edges of~$P_u$.
Then with high probability
\begin{equation}\label{e:slp}
 \sum_u R(P_u) \ge \sum_u \|u-r\|\ge c n n^{1/d}.
\end{equation}for some constant $c>0$. Indeed, with high probability, at least one half of the nodes lie at distance at least $\frac{1}{4}n^{1/d}$ from $r$.
Let $n_e$ be the number of paths $P_u$ that go through $e$, so
$n_e$ is the number of vertices below $e$ in the tree. Then
\[
 \sum R(P_u)=\sum_e R_e n_e.
\]
Indeed, $\sum R(P_u)$ counts the length of $e$ exactly $n_e$ times.
Now $\Ec_T=\sum R_e^\nu$, so by H\"older's inequality
\begin{equation}\label{e:E}
 \left(\sum \!\!R_e^{\nu}\right)^{1/\nu}\!\!\!
 \left(\sum n_e^{\nu/(\nu-1)}\right)^{(\nu-1)/\nu}\!\!\!\!\!\!\!\!\!
 \ge\sum_e R_e n_e\ge c n n^{1/d}.
\end{equation}
Thus it is enough to find an upper bound on $\sum n_e^{\nu/(\nu-1)}$.
If $e$ is at distance $i$ from $r$ then the latency of the tree
from $e$ onwards is at most $L^*+\delta-i$.
But this means it has at most $2^{L^*+\delta-i}\le (2n)2^{\delta-i}$
vertices. Hence $n_e\le n2^{1+\delta-i}$. Also, for each $i$ we have
\[
 \sum_{e:\dist(e,r)=i}n_e\le n
\]
as each vertex can be counted in at most one $n_e$ with $d(e,r)=i$.
Thus
\[
 \sum_{\dist(e,r)=i}n_e^{\nu/(\nu-1)}
 =\sum_{\dist(e,r)=i}n_e n_e^{1/(\nu-1)}\le n (n2^{1+\delta-i})^{1/(\nu-1)}
\]
for $i>\delta$, and
\[
 \sum_{\dist(e,r)=i}n_e^{\nu/(\nu-1)}\le n n^{1/(\nu-1)}
\]
for $i\le \delta$ as we always have $n_e\le n$.
The first sum is decreasing geometrically in~$i$, so
\[
 \sum_e n_e^{\nu/(\nu-1)}=(\delta+O(1))nn^{1/(\nu-1)}=O(1+\delta)n^{\nu/(\nu-1)}.
\]
Thus by \eqref{e:E},
\[
 \Ec_T^{1/\nu}(1+\delta)^{(\nu-1)/\nu}n\ge c'nn^{1/d}.
\]
Hence
\[
 \Ec_T=\Omega(n^{\nu/d}(1+\delta)^{1-\nu})
\]
as required.

The proof that $\pi^\Sum$ in Algorithm~\ref{alg:tradeoff} provides the correct upper bound is given in Appendix~\ref{proof:tradeoff}. 


\section{Proof of Theorem~\ref{t:clq}}\label{proof:clq}

For policy $\pi^\Gen$, the two stages of data forwarding and aggregation do not overlap. Hence, the latency and energy consumption under $\pi^\Gen$ are sum of the respective quantities in the two stages. The aggregation stage uses the  $\pi^\Sum$ policy and we have results from Theorem~\ref{t:main}. We now derive latency and energy scaling laws for the data-forwarding stage under $\pi^\Gen$ policy. Note that these depend on the function dependency graph $\Gmsc$.

We claim that the latency in the data forwarding stage of $\pi^\Gen$ is at most $ \Delta(\Gmsc)+1 $. This is because the forwarding graph is a subgraph of the dependency graph $(F^{\pi^\Gen}\subset \Gmsc)$ (with directions ignored) since each edge in $\Gmsc$ is traversed at most once, under the clique processor selection procedure described in Algorithm~\ref{alg:tradeoff_gen}. Note that  any graph with maximum degree $\Delta$ has a proper edge coloring using   at most $\Delta+1$ colors (Vizing's theorem) \cite{West:book2}. By scheduling transmissions of a single color in each timestep, we can ensure that each node is either transmitting/receiving from at most one node.  The energy consumption is at most the sum of the power-weighted edges of $\Gmsc$ and hence,
\[\Ec(\pi^\Gen;\delta,\Gmsc) \leq \sum_{e\in \Gmsc} R_e^\nu+ \Ec(\pi^\Sum;\delta,\emptyset).\] From \cite{Penrose&Yukich:03AAP}, for stabilizing graphs $\Gmsc$ (which include $k$-NNG and RGG), we have for uniform node sets $\bfV_n$,
\[ \lim_{n \to \infty} \frac{1}{n} \sum_{e\in \Gmsc(\bfV_n)} R_e^\nu\overset{L^2}{=} \zeta<\infty.\]
Hence, $\Ebb[\sum_{e\in \Gmsc} R_e^\nu] = \Theta(n)$ and we have the result.

\section{Upper bound construction for Theorem 1}\label{proof:tradeoff}

We shall need the following technical lemmas.

\begin{lemma}\label{l:t}
 Let\/ $C>1$, $d\ge 1$ and\/ $\nu \ge 0$ be constants.
 Let\/ $x$ be a point of the rectangular parallelepiped\/
 $Q=\prod_{i=1}^d [0,a_i]\subset \R^d$ of volume $n=\prod a_i$,
 and bounded aspect ratio, $a_i/a_j\le C$ for all\/ $i$,~$j$. Then
 as $n\to\infty$,
 \[
  \E(\min_{w\in\bfV_n}\|w-x\|^\nu)=\Theta(1),
 \]
 where the expectation is over all sets $\bfV_n$ of\/ $n$ points in
 $Q$ chosen independently and uniformly at random.

 The implied constants may depend on $C$, $d$ and\/ $\nu$,
 but not on~$n$.
\end{lemma}
\bprf
Let $r>0$ and set $\Vmsc_r$ to be the volume of the set of points in $Q$
that are within distance $r$ of $x$. Clearly $\Vmsc_r$ is at most the
volume of a $d$-dimensional sphere of radius~$r$, i.e., $\Vmsc_r\le cr^d$
for some constant $c=c(d)$. For a lower bound we note that
$\Vmsc_r\ge 2^{-d}c r^d$ if $r$ is sufficiently small, the worst case
being when $x$ is at a corner of~$Q$. However, for a general bound
valid for larger $r$ we use the following observation.
If $r=\diam Q$, the diameter of $Q$, then
$\Vmsc_r$ is the volume of $Q$, which is just~$n$. If $r\le \diam Q$
then, by shrinking $Q$ by a linear factor $r/\diam Q$ about $x$
and noting that the resulting set $Q'$ lies inside $Q\cap \Vmsc_r$, we see
that $\Vmsc_r\ge |Q'|=(r/\diam Q)^d n$. As $a_i/a_j\le C$ for all $i,j$,
we have $\diam Q=\Theta(n^{1/d})$ and hence $\Vmsc_r\ge c'r^d$
for some constant $c'=c'(d,C)>0$.

Using integration by parts we can write
\begin{align*}
 \E(\min_{w\in\bfV_n}\|w-x\|^\nu)
 &=\int_0^{\diam Q}r^\nu(-\tfrac{d}{dr}\Prb(\min_{w\in\bfV_n}\|w-x\|\ge r)\,dr,\\
 &=\big[-r^\nu\Prb(\min_{w\in\bfV_n}\|w-x\|\ge r)\big]_0^{\diam Q}\\
 &\quad +\int_0^{\diam Q}\big(\tfrac{d}{dr}r^\nu\big)\Prb(\min_{w\in\bfV_n}\|w-x\|\ge r)\,dr\\
 &=\int_0^{\diam Q}\nu r^{\nu-1}\Prb(\min_{w\in\bfV_n}\|w-x\|\ge r)\,dr
\end{align*}
as $\nu>0$ and (for $n>0$) there is always a point within distance $\diam Q$ of~$x$.
However, $\min_{w\in\bfV_n}\|w-x\|\ge r$ if and only if there is no point of $\bfV_n$
in~$\Vmsc_r$. Thus
\[
 1-\Vmsc_r\le \Prb(\min_{w\in\bfV_n}\|w-x\|\ge r)
 =(1-\Vmsc_r/n)^{n}\le \exp(-\Vmsc_r).
\]
Hence
\[
 \int_0^{1/c^{1/d}}\nu r^{\nu-1}(1-cr^d)\,dr\le
 \E(\min_{w\in\bfV_n}\|w-x\|^\nu)
 \le \int_0^{\infty}\nu r^{\nu-1}\exp(-c'r^d)\,dr,
\]
where we have restricted the range of integration
to a constant for the lower bound (which is less than $\diam Q=\Theta(n^{1/d})$
for sufficiently large~$n$), and extended the range on the upper bound.
However, both bounds are now positive constants independent of~$n$, although
they do depend on $d$, $\nu$, and~$C$.
Thus $\E(\min_w\|w-x\|^\nu)=\Theta(1)$ as required.
\eprf

Given a set $\bfV_n\subseteq\R^d$ and two points $u,v\in\R^d$,
a $\bfV_n$-path of length $k$ from $u$ to $v$ is a sequence
$ux_1x_2\dots x_{k-1}v$ with $u_1,\dots,u_{k-1}\in\bfV_n$. As before,
the {\em energy} of a path $P=u_0u_1\dots u_k$ is
\[
 E_P=\sum_{i=1}^k \|u_i-u_{i-1}\|^\nu.
\]

\begin{lemma}\label{l:path}
 Let\/ $u$ and $w$ be  points in $Q=[0,n^{1/d}]^d$ at distance
 $R_{uw}=\|u-w\|$ from each other and let\/ $\bfV_n$ be a set of\/ $n$
 points in $Q$ chosen uniformly and independently. Let\/ $E_k(u,v;\bfV_n)$ be
 the minimal energy of a $\bfV_n$-path of length at most\/ $k$ from $u$ to~$v$.
 Then
 \[
  \E\big(E_k(u,v;\bfV_n)\big)=O\big(\max\{k(R_{uw}/k)^\nu,R_{uw}+1\}\big).
 \]
\end{lemma}
\bprf
Let $x_1,\dots,x_{k-1}$ be $k-1$ subdivision points on the line segment
from $u$ to~$v$, equally spaced so that this line segment is divided into
$k$ equal segments. Pick $u_i$ to be the point of $\bfV_n$ closest to $x_i$,
and also write $u_0=x_0=u$ and $u_k=x_k=v$. Then, after possibly removing loops
and repeated vertices, $u_0u_1\dots u_k$ gives a suitable path from $u$ to $v$
of energy at most $\sum_{i=1}^k \|u_i-u_{i-1}\|^\nu$. Now
\begin{align*}
 \|u_i-u_{i-1}\|^\nu
 &\le 3^\nu(\tfrac13(\|u_i-x_i\|+\|u_{i-1}-x_{i-1}\|+\|x_i-x_{i-1}\|))^\nu\\
 &\le 3^\nu\max\{\|u_i-x_i\|^\nu,\|u_{i-1}-x_{i-1}\|^\nu,\|x_i-x_{i-1}\|^\nu\}.
\end{align*}
Hence the energy of the path is at most $O(k(R_{uw}/k)^\nu)+O(\sum\|u_i-P\|^\nu)$.
By \Lm{t}, $\E(\sum\|u_i-x_i\|^\nu)=k\Theta(1)$. If $k\le R_{uw}$
then the expected energy of the path is at most $O(k(R_{uw}/k)^\nu)$, as
required. If $k>\ell_{uw}$, then we perform the same construction replacing $k$ with
$\lceil R_{uw}\rceil$ to obtain a path of expected weight $O(R_{uw}+1)$.
\eprf

Given $\nu$, $d$, $\delta$, our aim is construct a spanning tree $T$ with latency
at most $L^*(n)+\delta$ and small energy $E_T=\sum R_e^\nu$. The tree
will be formed from a minimal latency tree $T_L$ with latency
$L=L^*(n)=\lceil\log_2n\rceil$. Some edges will be subdivided and not
all vertices will be present. Some other minor changes to the tree may
also be made. The edges of $T_L$ of level $i$ will be subdivided at most
\begin{equation}\label{e:si}
 s_i:= \lfloor c_1\delta 2^{i(1/\nu-1/d)}\rfloor
\end{equation}
times when $\nu\ge d$, where $c_1=1-2^{1/\nu-1/d}$ (or
$c_1=1/\lceil\log_2n\rceil$ if $\nu=d$) is a normalizing factor chosen so
that $\sum_i s_i\le \delta$. For $\nu<d$ we shall take $s_i=0$ as subdividing
is not necessary. We will of course need to prune some leaves later so that
the tree has $n$ vertices. It is clear that the latency of $T_L$ with level $i$
edges subdivided at most $s_i$ times is at most
$L^*+\sum s_i\le L^*(n)+\delta$.
Indeed we just replace the single time step $L-i$ with $1+s_i$ time steps,
during which data is transmitted along the (at most) $1+s_i$ edges of each
subdivided level $i$ edge of~$T_L$.

Our strategy is as follows. Given a rectangular $d$ dimensional region $Q$,
containing $n$ points, one of which is chosen as a root vertex $v$, we pick a
coordinate, say $x_1$, in which $Q$ has largest extent and subdivide
$Q=Q_0\cup Q_1$ so that half of the points are on each side.

The root vertex $v$ will be in one half, say $Q_0$. Choose a minimal
weight path $P$ from $v$ with at most $s_0$ subdivisions joining $v$
to a vertex $v_1$ in $Q_1$. Now inductively repeat the construction
within $Q_0$ and $Q_1$ using $v_0=v$ and $v_1$ as the corresponding
root vertices. At the next stage, each of $Q_0$ and $Q_1$ is subdivided
as above, say $Q_0=Q_{00}\cup Q_{01}$ and $Q_1=Q_{10}\cup Q_{11}$.
Assume w.l.o.g.\ that $v_0\in Q_{00}$, $v_1\in Q_{10}$. Pick a vertex
$v_{01}\in Q_{01}$ that has not already been used and join with a
path with at most $s_1$ subdivisions to $v_{00}=v_0$ inside $Q_0$.
Similarly join some $v_{11}\in Q_{11}$ to $v_{10}=v_1$ inside $Q_1$.
Repeat this process. At each stage, choose the $v_{\dots 1}$
to be any vertex in the other half of the appropriate region that
has not already been used on any path so far. If no such vertex exists then
the construction within this half terminates as we have exhausted
all the vertices.

There are two main problems with this approach, both concerning
the applicability of induction to the process. The $n$ points in $Q$
were chosen to be {\em uniform} and {\em independent}.
If care is not taken with the division of $Q$ into $Q_0$ and $Q_1$
then the points in $Q_0$ and $Q_1$ will not be uniform. For
example, it is {\em not} sufficient to order the points by the
$x_1$-coordinate and just take some subdivision
between the $(n/2)$th and $(n/2+1)$st points. Neither
will the volumes of $Q_0$ and $Q_1$ be exactly $n/2$, although this
is a relatively minor problem as we shall see later. The second, and
more serious problem is that after constructing the path from
$v_0$ to $v_1$, some vertices in $Q_0$ and $Q_1$ have now been
used. the remaining points are now {\em not\/} independent. This second
point will be dealt with by allowing the subdividing vertices
to be reused in subsequent steps, i.e., by allowing subdividing
paths to intersect. The resulting graph will then not be a tree,
but we shall show that it will still have the same latency
and can be modified to form a tree with no increase in latency.
To solve the first problem mentioned above we shall use the following lemma.

\begin{lemma}\label{l:beta}
 Take $n$ random points independently and uniformly in $[0,1]$
 and order them by value as $0\le x_1\le x_2\le \dots\le x_n\le 1$.
 Then $x_{r+1}$ has mean $\frac{r+1}{n+1}$ and variance
 $\frac{(r+1)(n-r)}{(n+1)^2(n+2)}$. Moreover,
 conditioned on the value of $x_{r+1}$, the set of values
 $\{x_1,\dots,x_r\}$ can be obtained by taking $r$ random
 points independently and uniformly distributed in $[0,x_{r+1}]$.
\end{lemma}
\bprf
The probability that $x_{r+1}$ lies in a small interval $[x,x+dx]$
is given as the probability that {\em some} point is in the
interval times the probability that there are exactly $r$ of
the remaining $n-1$ points in $[0,x]$. Thus the pdf of $x_{r+1}$
is $n\binom{n-1}{r}x^r(1-x)^{n-r-1}$. This is just the pdf
of the beta distribution with parameters $r+1$ and $n-r$. The
mean and variance can be obtained by using standard results on the beta
distribution, or by a straightforward calculation. For the
conditioning result, imagine the points are located in small bins
of size $dx$, chosen so that it is very unlikely that two points
appear in the same bin. Conditioning on $x_{r+1}$ lying in the
bin $[x,x+dx]$ is equivalent to conditioning on one point
lying in this interval, $r$ points lying in earlier bins, and
$n-r-1$ points lying in later bins. However and valid assignment
of points to bins is equally likely, so the conditional distribution
of the first $r$ points is equivalent to choosing $r$ points
uniformly and independently in $[0,x]$.
\eprf

\begin{corollary}\label{c:emax}
 If\/ $r=n/2+O(1)$ in \Lm{beta} and $c=\frac{1}{2}+O(1/n)$, then for $\nu>0$
 \[
   \E(\max\{x_{r+1}/c,1\}^\nu)=1+O(1/\sqrt{n}).
 \]
\end{corollary}
\bprf
Note that $x_{r+1}/c$ is bounded between 0 and $2+\varepsilon$ for
large $n$, so $\max\{x_{r+1}/c,1\}^\nu\le 1+K|x_{r+1}-c|$ for some
constant $K=K(\nu)>0$. Now
\[
 \E(|x_{r+1}-c|)^2\le \E(|x_{r+1}-c|^2)=\Var(x_{r+1})+(\E(x_{r+1})-c)^2.
\]
However, by \Lm{beta}, $\E(x_{r+1})-c=O(1/n)$ and $\Var(x_{r+1})=O(1/n)$.
The result follows.
\eprf\\


\noindent{\em Proof of the upper bound for Theorem 1: }
We follow the strategy outlined above. At step $i$ have
$n_i$ vertices, $\lfloor n/2^i\rfloor\le n_i\le\lceil n/2^i\rceil$,
chosen uniformly at random from a rectangular parallelepiped
$Q=\prod [0,a_j]$ of volume $n_i$ and aspect ratio $a_j/a_k\le 3$
for all $j,k$. We wish to construct a spanning tree of latency at most
$L_i=L^*(n)-i+\sum_{j\ge i}s_j$ of expected energy
at most $M_i$; the value of $M_i$ to be determined.

If $n_i=1$ then there is nothing to do, so assume $n_i>1$.
Pick a coordinate $x_j$ with maximum value of $a_j$. W.l.o.g.\ $j=1$.
Order the $n_i$ points by their $x_1$-coordinates, which are uniform
and independent in $[0,a_1]$.
Let $x_1(r)$ be the value of the $r$th $x_1$-coordinate.
Define $Q_0$ to be the subset of points of $Q$ with $x_1$-coordinate
less than $x_1(n_{i+1}+1)$, where $n_{i+1}=\lfloor n_i/2\rfloor
\in[\lfloor n/2^{i+1}\rfloor,\lceil n/2^{i+1}\rceil]$.
Let $Q_1$ to be the subset of points of $Q$ with $x_1$-coordinate
more than $x_1(n_{i+1})$. Note that $Q_0$ and $Q_1$ overlap,
however their intersection contains no points of $\bfV_n$.
Also note that (with probability 1), there are $n_{i+1}$
points in $Q_0$ and $n_i-n_{i+1}=\lceil n_i/2\rceil
\in [\lfloor n/2^{i+1}\rfloor,\lceil n/2^{i+1}\rceil]$ points in $Q_1$.
By \Lm{beta}, conditioned on the shape of $Q_0$, say, the points in $Q_0$
are uniformly distributed inside $Q_0$. (There is however
a dependency of these points on the shape of $Q_1$.) If the root lies
in $Q_0$ then we use this as the root of $Q_0$, otherwise
we pick any point in $\bfV_n\cap Q_0$ as the root. Shrink $Q_0$
in the $x_1$-direction by a factor
$X_0=(x_1(n_{i+1}+1)/a_1)/(n_{i+1}/n_{i})$
to obtain a set $Q'_0$ with volume exactly $n_{i+1}$.
Now construct a spanning tree of latency at most
$L_{i+1}$ on the points in $Q'_0$. Let $E_0$ be the energy of this tree.
Now scaling by a factor of $X_0$ we obtain a tree $T_0$ in $Q_0$ of
expected energy at most $\max\{1,X_0\}^\nu E_0$, all lengths
having increased by a factor of at most $\max\{1,X_0\}$. Note that
$X_0$ and $E_0$ are independent, so the expected value of the energy
of this tree is at most $\E(\max\{1,X_0\}^\nu)\E(E_0)$ which
is at most $(1+C/\sqrt{n_i})\E(E_0)$ by \Co{emax}.
Since $a_1$ was maximal and the $x_1$-extent of $Q'_0$ is
$a_1 n_{i+1}/n_i\ge a_1/3$, the aspect ratios of $Q'_0$ are again at
most~3. The points in $Q_0$ are distributed uniformly at random and
the volume of $Q'_0$ is $n_{i+1}$. Thus by reverse induction on $i$,
$\E(E_0)\le M_{i+1}$. A similar argument also applies to $Q_1$,
so we obtain two trees with expected total energy of at most
$(1+C/\sqrt{n_i})M_{i+1}$. By \Lm{path} we can find a path
from the root $v_0$ of $Q_0$ to the root $v_1$ of $Q_1$ of length
at most $1+s_i$ and expected energy at most
\[
 C'\max\{n_i^{\nu/d}(1+s_i)^{1-\nu},n_i^{1/d}+1\big\}
\]
Note that this path will reuse vertices of the trees $T_0$ and $T_1$,
so that we may obtain a graph $G$ with cycles. However, the latency of $G$
is still at most $L_i$. Indeed, in the first
$L_{i+1}=L^*(n)-(i+1)+\sum_{j\ge i+1}s_j$
steps we transmit the data to $v_0$ and $v_1$ in $T_0$ and $T_1$
separately. Then in the last $1+s_i$ steps we transmit the data
along the $v_0-v_1$ path $P$ to whichever of these vertices was the root
of~$Q$. If one insists on having a tree then one can modify $G$
slightly as follows. Any vertex of $T_0$ or $T_1$ that lies in $P$
holds the data rather than transmitting it on during the first
$L_{i+1}$ steps. Tree edges from these vertices
towards the roots $v_0$ and $v_1$ are deleted from $T_0$ and $T_1$
as these edges are no longer used. The trees $T_0$ and $T_1$ now
become a forest of possibly many trees. All the data after the first
$L_{i+1}$ steps is located at vertices that are roots of all the individual
trees in this forest. These vertices all lie on $P$. The path $P$ then joins
all the roots of these small trees to form one tree and data
can be swept up along $P$ in the last $1+s_i$ time steps.

It now remains to bound the expected energy of this tree.
We have an upper bound given by the energies of $T_0$, $T_1$, and $P$
of
\begin{equation}\label{e:mi}
 M_i:=2(1+C/\sqrt{\tilde n_i})M_{i+1}+C'\max\{n_i^{\nu/d}(1+s_i)^{1-\nu},n_i^{1/d}+1\}
\end{equation}
where in order to maximize the expression we take $\tilde n_i=\lfloor n/2^i\rfloor$
and $n_i=\lceil n/2^i\rceil$. Note $M_i=0$ for $i\ge \log_2 n$ as then $n_i\le 1$.
Inductively substituting the bound in \eqref{e:mi} for $M_{i+1}$ gives
\[
 M_i\le\Big(\prod_{j>i}(1+C/\sqrt{\tilde n_j})\Big)
 \sum_{j\ge i}C'2^{j-i}\max\{n_j^{\nu/d}(1+s_j)^{1-\nu},n_j^{1/d}+1\}
\]
Now the $\tilde n_j$ increase exponentially as $i$ decreases from $\log_2 n$,
thus the product $\prod_{j>i}(1+C/\sqrt{\tilde n_j})$ can be bounded independently
of $n$ and~$i$. Hence
\begin{align}
 M_0&\le C''\sum_{i=0}^{\log_2n}2^i\max\{(n/2^i)^{\nu/d}(1+s_i)^{1-\nu},(n/2^i)^{1/d}+1\}\notag\\
 &\le C''\sum_{i=0}^{\log_2n}2^i(n/2^i)^{\nu/d}(1+s_i)^{1-\nu}
 +C''\sum_{i=0}^{\log_2n}2^i((n/2^i)^{1/d}+1)\label{e:m0}
\end{align}
where $C''$ absorbs this product and also any factors that arise from the difference
between $n_j$ and $n/2^j$. The second sum in \eqref{e:m0} is geometrically
increasing for $d>1$ so has sum $O(n)$. Assume $\nu>d$.
Then $s_i=\Theta(\delta 2^{i(1/\nu-1/d)})$ and
$1+(1/\nu-1/d)(1-\nu)-\nu/d=(1/\nu-1/d)<0$. Thus the first sum in \eqref{e:m0}
is decreasing geometrically when $s_i>0$, and also decreasing geometrically
at an even faster rate when $s_i=0$. Hence this sum gives
$O(n^{\nu/d}(1+s_0)^{1-\nu})$. Thus
\[
 M_0=O(\max\{n,n^{\nu/d}(1+\delta)^{1-\nu}\}).
\]
for $\nu>d$. For $\nu=d>1$, $s_i=s_0=\Theta(\delta/\log n)$ is constant.
Thus the terms in the first sum are equal and add up to
$(\log n)O(n(1+s_0)^{1-\nu})$. Thus
\[
 M_0=O(\max\{n,n(\log n)(1+\delta/\log n)^{1-\nu}\}).
\]
for $\nu=d>1$. Finally, for $\nu<d$, $s_i=0$ and the first sum is increasing geometrically.
This sum is then $O(n)$ and so
\[
 M_0=O(n)
\]
for $1<\nu<d$. \qed

\end{appendix}



\end{document}